\documentclass{iopart}
\usepackage{graphicx}
\newcommand{\beq}{\begin{equation}}
\newcommand{\eeq}{\end{equation}}
\newcommand{\beqar}{\begin{eqnarray}}
\newcommand{\eeqar}{\end{eqnarray}}
\newcommand{\beqars}{\begin{eqnarray*}}
\newcommand{\eeqars}{\end{eqnarray*}}
\newcommand{\bc}{\begin{center}}
\newcommand{\ec}{\end{center}}
\newcommand{\ben}{\begin{enumerate}}
\newcommand{\een}{\end{enumerate}}
\newcommand{\bit}{\begin{itemize}}
\newcommand{\eit}{\end{itemize}}

\def \veps{\varepsilon}

\def\f{{\bf f}-}

\def\r{{\bf r}-}
\def\g{{\bf g}-}
\def\1{{\bf g$_1$}-}
\def\2{{\bf g$_2$}-}

\renewcommand{\(}{\left(}
\renewcommand{\)}{\right)}

\begin{document}
\jl{6}

\title[Gravitational Waves from Rotating Proto-Neutron Stars]
{Gravitational Waves from Rotating Proto-Neutron Stars}

\author{V.Ferrari\ddag, L.Gualtieri\ddag, J.A.Pons\S, A.Stavridis\ddag}
\address{\ddag\ Dipartimento di Fisica ''G. Marconi'', Universit\`a di Roma
''La Sapienza'' and Sezione INFN ROMA 1, piazzale Aldo Moro 2, I--00185 Roma,
Italy}
\address{\S\ Departament d'Astronomia i Astrof\'{\i}sica, Universitat de 
Val\`encia, 46100 Burjassot, Val\`encia, Spain}

\begin{abstract}
  We study the effects of rotation on the quasi normal modes (QNMs) of
  a newly born proto neutron star (PNS) at different evolutionary
  stages, until it becomes a cold neutron star (NS).  We use the
  Cowling approximation, neglecting spacetime perturbations, and
  consider different models of evolving PNS.  The frequencies of the
  modes of a PNS are considerably lower than those of a cold NS, and
  are further lowered by rotation; consequently, if QNMs were 
  excited in a sufficiently energetic process, they would radiate
  waves that could be more easily detectable by resonant-mass and
  interferometric detectors than those emitted by a cold NS.  We find
  that for high rotation rates, some of the \g-modes become unstable
  via the CFS instability; however, this instability is likely to be
  suppressed by competing mechanisms before emitting a significant
  amount of gravitational waves.
\end{abstract}
\label{firstpage}
\section{Introduction}
It has recently been shown \cite{FMP} that, during the first minute of
life of a proto-neutron star (PNS) born in a gravitational collapse,
the frequencies of its quasi normal modes (QNMs) change. Indeed,
the star cools down and contracts and its internal structure is modified
by neutrino diffusion and thermalisation processes.
The evolutionary models used in \cite{FMP}, developed in \cite{Pons1,Pons2},
describe the stellar evolution in terms of a sequence of
equilibrium configurations; this quasi-stationary description 
has been shown to become appropriate 
after a few tenths of seconds from the stellar birth.
By solving the equations of stellar perturbations it has been shown that the
frequencies of all QNMs of a newly born PNS are much smaller than
those of the cold NS which forms at the end of the evolution.
In order to isolate the effects of the thermal and chemical processes
from those induced by rotation,  in \cite{FMP} all stellar models were 
assumed to be non rotating.
Subsequently, this study has been generalized 
to include rotation \cite{nostro}, and to 
explore the possibility that the modes  become unstable due to
Chandrasekhar--Friedman--Schutz (CFS) mechanism
\cite{Chandra,FriedSchutzA,FriedSchutzB}.

The CFS instability of the fundamental mode (\f-mode) has been studied
extensively in the literature \cite{newt1}-\cite{yo3}, and it has been
shown to act at very high stellar rotation rates, comparable to the
break-up limit. Furthermore, unless the temperature is very low
viscous dissipation mechanisms tend to stabilise the \f-mode
instability.  The \r-mode instability has also been extensively
studied in recent years, after Andersson \cite{Nils} discovered that
it is generic for every rotating star. However, whether or not this
instability removes a considerable amount of rotational energy from
the star is still matter of debate.

Since in the no rotation limit the \g-modes have frequencies lower
than that of the \f-mode, they may become unstable for relatively
small values of the angular velocity.  This provides a good motivation
for the study of CFS instability of \g modes, and indeed it   has
been studied in \cite{Lai} for zero temperature stars using the
Newtonian theory of stellar perturbations in the Cowling
approximation. This study regarded a particular class of \g-modes for
which the buoyancy, which provides the restoring force for the modes,
is due to the gradient of proton to neutron ratio in the interior of the star.

In \cite{nostro} we studied the onset of the CFS
instability of the lowest \g-modes of a newly born, hot proto-neutron
star, where the buoyancy is mainly due to the high entropy and composition
gradients that dominate the stellar dynamics.
We used the relativistic theory of stellar perturbations of a
slowly rotating star in the Cowling approximation, which is known to
reproduce with a good accuracy the \g-mode frequencies, because such  
modes are associated with small gravitational perturbations.

We stress that today it is not definitely clear if CFS instability is
relevant in newly born neutron stars. This and other similar
mechanisms could explain why we do not observe pulsars spinning with
periods of less than 5ms, which are predicted by most stellar
models. \cite{nature}
\section{Formulation of the problem}
We consider a relativistic star in uniform rotation with an angular
velocity $\Omega$ so slow that the distortion of its figure from
spherical symmetry is of order $\Omega^2$, and can be ignored.
Following the approach of Hartle \cite{hartle}, we expand all
equations with respect to the parameter $\veps = \Omega / \Omega_K$,
where $\Omega_K=\sqrt{\frac{M}{R^3}}$; we retain only first order
terms ${\cal O}( \veps )$.  On these assumption, the metric can be
written in the form
\begin{equation}
  \label{metric}
  ds^2 = -e^{2\nu(r)}dt^2
  + e^{2\lambda(r)} dr^2 + r^2\( d\theta^2  + \sin^2\theta d\phi^2\)
  - 2\veps \omega(r)\sin^2\theta dt d\phi.
\end{equation}
The star is assumed to be composed by a perfect fluid, whose 
energy momentum tensor is
\begin{equation}
  \label{en_mom_tensor}
  T_{\mu\nu} = \(p + \rho\) u_{\mu}u_{\nu} + pg_{\mu\nu},
\end{equation}
with pressure $p$, energy density $\rho$ and four-velocity components
$ u^{\mu} = [ e^{-\nu},0,0,\Omega e^{-\nu} ].$ The metric functions
$\nu(r),\lambda(r)$ are found by solving the equations of hydrostatic
equilibrium \cite{hartle}.  We use as a background the models of
evolving proto-neutron stars developed in \cite{Pons1,Pons2}, and
study the non--radial, adiabatic perturbations of these models for
selected values of the evolution time; we start from $t = 0.5~$s after
the formation of the proto-neutron star, when the quasi--stationary
description becomes appropriate to represent the stellar evolution.

The complete set of equations for the perturbations has been derived
using the so--called BCL gauge \cite{BCL} by Ruoff, Stavridis \&
Kokkotas \cite{RSK1}. The Cowling limit of these equations, which
neglects the contribution of the gravitational perturbations, was
studied in \cite{RSK2} for polytropic relativistic equations of state.
In \cite{nostro} we have used the Cowling approximation and, since we
are interested in the evolution of the \g and \f modes, that have polar
parity, we have neglected the coupling with axial parity
perturbations.  The results we show refer to the components with
harmonic indices $l=m=2$, which are the most relevant for
gravitational wave emission.
\section{Results}
In order to find the frequencies of the quasi--normal modes, we have
numerically integrated the perturbed equations both in the time and in
the frequency domain for different values of the evolution time
$t_{ev}$, and for selected values of the rotation parameter
$\epsilon$. The rotation rate has been chosen to vary within
$0\le\epsilon\le 0.4$ because from preliminary calculations we find
that for the models under consideration the mass shedding limit does
not exceed $\veps=0.4$.  The results refer to the evolutionary model
labelled as model A in \cite{FMP}.  We consider the first minute of
life of the proto-neutron star, from $ t_{ev} = 0.5~ s$ to $ t_{ev} =
40 ~s$, during which the star significantly cools down and contracts,
and processes related to neutrino diffusion and thermalization
dominate the stellar evolution.  The gravitational mass of the star,
which is $M=1.56~M_\odot$ at $t_{ev}=0.2~s$, due to neutrino emission
becomes $M=1.46~M_\odot$ at $t_{ev}=40~s$.  The radius of the initial
configuration is $R=23.7$ km and reduces to $R=12.8$ km at
$t_{ev}=40~s$.

\begin{figure}
\begin{center}
\includegraphics[width=5.5cm,height=4.5cm,angle=-90]{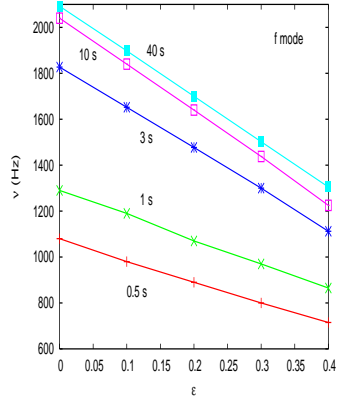}
\caption{
The frequency of  the fundamental  mode  of  a newly born PNS
is plotted as a function  of the rotational parameter $\veps = \Omega / \Omega_K,$
for assigned values of  the time elapsed from the  stellar birth.
As the time grows, the frequency increases and tends
to that of the cold neutron star which forms at the end of the evolutionary
process. The onset of the CFS instability occurs when the 
mode frequency becomes zero, and we see that the \f-mode would
become unstable only for extremely high values of the rotational parameter, 
 as it is for cold stars.
}
\label{fig1}
\end{center}
\end{figure}

\begin{figure}
\begin{center}
\includegraphics[width=5.5cm,height=4.5cm,angle=-90]{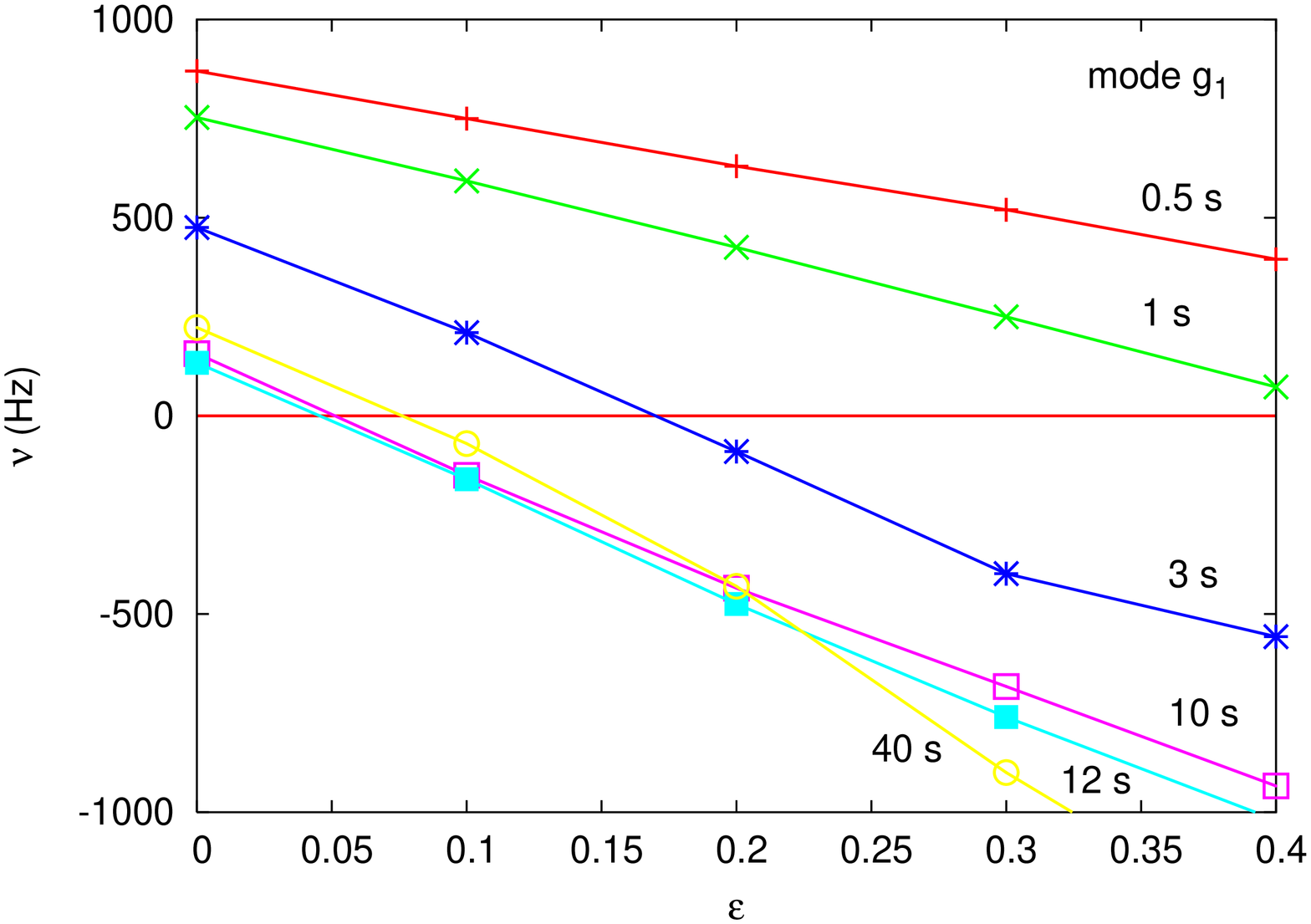}
\includegraphics[width=5.5cm,height=4.5cm,angle=-90]{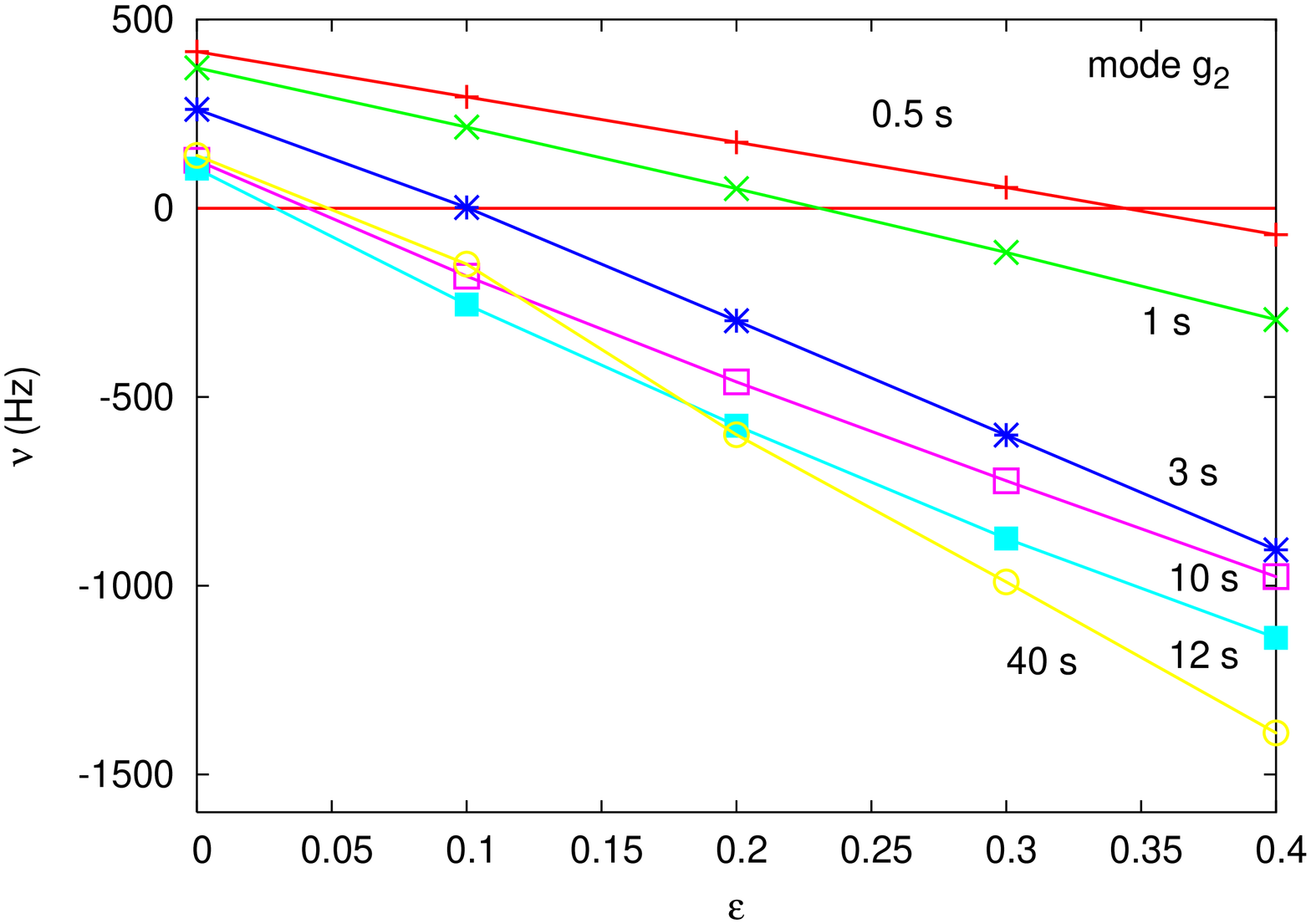}
\caption{
The frequency of  the \1 and \2  modes  of a newly born  PNS are
plotted, as in figure 1, for the same stellar models.
Unlike the \f-mode, as the time grows the frequency  of the \g-modes
decreases, reaches a minimum  at about $t_{ev}=12$ and
then slightly increases (see text). 
For both modes  the CFS instability sets in at values of the
rotational parameter much lower than that needed for the \f-mode.
}
\label{fig2}
\end{center}
\end{figure}

The main results of our work are summarized in figures \ref{fig1}
and \ref{fig2}, where we plot the frequencies of the \f, \1 and \2 modes as a 
function of the rotation parameter $\veps= \Omega / \Omega_K$,
for different values of the evolution time in the more interesting phases 
of the cooling process.
We see that as the time elapsed from the birth of the PNS grows, 
the \f-mode frequency increases and tends
to that of the cold neutron star which forms at the end of the evolutionary
process. Conversely, the frequency  of the \g-modes initially decreases, reaches
a minimum  for $t_{ev} \simeq 12$ s, and subsequently smoothly increases.
This behaviour is justified by the fact that
the \g modes are associated to entropy and composition gradients, and
whereas during the first 10-12 seconds the dynamical evolution of
the star is dominated by strong entropy gradients that progressively
smooth out, after  $\sim 12~s$ the entropy becomes nearly constant
throughout the star and {\bf g}-modes due to composition gradients
take over.

The onset of the CFS instability is
signaled by the vanishing of the mode frequency for some value of the
angular velocity (neutral point).  From figures \ref{fig1} and
\ref{fig2} we see that while the \f-mode does not become unstable
during the first minute of the PNS life, both the \1 and the \2 modes
do become unstable.  The \1 frequency remains positive during the
first second, but at later times it vanishes for very low values of
$\veps.$ For instance, at $t_{ev}=3 ~s$ it crosses the zero axis for
$\Omega =0.17~\Omega_K,$ even though its value for the corresponding
nonrotating star is still quite high, $\nu_{g_1}=486 ~$Hz.  The
behaviour of the \2 mode is similar, but being the frequency lower the
instability sets in at lower rotation rates.  

It should be mentioned that in \cite{FMP} we studied also a second
model of evolving proto-neutron star, labelled as model B.  The main
difference between the two models is that model A has an equation of
state softer than model B, and that at some point of the evolution a
quark core forms in the interior of model B.  We have integrated the
perturbed equations also for model B, finding results entirely similar
to those of model A; this indicates that the quark core that develops
at some point of the evolution does not affect the overall properties
of the modes in a relevant way.

A mode instability is physically significant if its growth time is
sufficiently small with respect to the timescales typical of the
stellar dynamics; in this case the instability has sufficient time to
grow before other processes damp it out or the structure of the
evolving star changes.  In \cite{nostro} we give an ``order of
magnitude" estimate of the growth time of the \g-modes as follows: we
compute the energy $E$ associated with a given mode in Newtonian
approximation as in \cite{Lai,FriedSchutzA}, and the gravitational
luminosity $\frac{dE}{dt}$ using a multipole expansion as in
\cite{Lind}.  The growth time associated to gravitational radiation
reaction then is \beq {1 \over \tau_{gr}} = - {1 \over 2E} {dE \over
  dt}.  \eeq Since we are working in the Cowling approximation, we
neglect the perturbation of the Newtonian potential. Furthermore, we
neglect the contribution due to current multipoles, because they
correspond to axial parity perturbations.

The growth times of the unstable \1 modes shown in figure \ref{fig2}
are summarized in Table 1.  The growth time appears to be orders of
magnitude larger than the evolutionary timescale, which is of the
order of tens of seconds.  Although the estimate based on Newtonian
expressions is a quite crude one, the growth time is so much larger
than the evolutionary timescale that it is reasonable to conclude that
the CFS instability of the lowest \g-mode is unlikely to play a
relevant role in the early evolution of proto-neutron stars.  Similar
conclusions can be drawn for the fundamental mode and for higher order
{\bf g}-modes.
\section{Aknowledgements}
We would like to thank G. Miniutti for suggesting that the \g-modes of
newly born PNSs may be subjected to the CFS instability.

This work has been supported by the EU Program 'Improving the Human Research 
Potential and the Socio-Economic Knowledge Base' 
( Research Training Network Contract HPRN-CT-2000-00137 ).

\label{lastpage}

\begin{table}
\centering
\caption{Growth times for unstable $g_1$ mode of Model A for 
$t_{ev}=3~s,\,12~s,\,40~s$}
\begin{tabular}{*{3}{c}}
\hline\hline
$\veps$ & $\nu$~(Hz) & $\tau_{gr}~ (s) $ \\[0.5ex]
\hline\hline
0.1 & 200 & \dots \\[0.5ex]
0.2 & -90 & -1.5 $10^{9}$  \\[0.5ex]
0.3 & -683 & -2 $10^6$  \\[0.5ex]
0.4 & -789 & -2.2 $10^{4}$  \\[0.5ex]
\hline\hline
\label{modelAt3}
\end{tabular}
\begin{tabular}{*{3}{c}}
\hline\hline
$\veps$ & $\nu$~(Hz) & $\tau_{gr}~ (s) $ \\[0.5ex]
\hline\hline
0.1 & -160 & -4.6 $10^{7}$  \\[0.5ex]
0.2 & -475 & -1.7 $10^{6}$  \\[0.5ex]
0.3 & -760 & -3.4 $10^5$  \\[0.5ex]
0.4 & -1020 & -2.7 $10^{5}$  \\[0.5ex]
\hline\hline
\label{modelAt12}
\end{tabular}
\begin{tabular}{*{3}{c}}
\hline\hline
$\veps$ & $\nu$~(Hz) & $\tau_{gr}~ (s) $ \\[0.5ex]
\hline\hline
0.1 & -70 & -3.7 $10^{10}$  \\[0.5ex]
0.2 & -430 & -1.3  $10^{9}$ \\[0.5ex]
0.3 & -900 & -3.6 $10^6$  \\[0.5ex]
0.4 & -1250 & -1.3 $10^5$  \\[0.5ex]
\hline\hline
\label{modelAt40}
\end{tabular}
\end{table}

\end{document}